\documentclass[12pt]{article}
\usepackage{graphicx}

\newcommand\pubdate{\today}

\def\napoli{Institute of Nuclear Physics Polish Academy of Sciences, PL-31342 Krakow, Poland}

\def\Title#1{\begin{center} {\Large #1 } \end{center}}
\def\Author#1{\begin{center}{ \sc #1} \end{center}}
\def\Address#1{\begin{center}{ \it #1} \end{center}}

\newcommand\pubblock{\rightline{\begin{tabular}{l} \\%\pubnumber\\
         \pubdate  \end{tabular}}}
\newenvironment{Abstract}{\begin{quotation}  }{\end{quotation}}
\newenvironment{Presented}{\begin{quotation} \begin{center} 
             PRESENTED AT\end{center}\bigskip 
      \begin{center}\begin{large}}{\end{large}\end{center} \end{quotation}}
\def\Acknowledgements{\bigskip  \bigskip \begin{center} \begin{large}
             \bf ACKNOWLEDGEMENTS \end{large}\end{center}}
\def\beq{\begin{equation}}
\def\eeq#1{\label{#1}\end{equation}}
\def\eeqn{\end{equation}}

\def\beqa{\begin{eqnarray}}
\def\eeqa#1{\label{#1}\end{eqnarray}}
\def\eeqan{\end{eqnarray}}

\let\bar=\overbar

\def\Dslash{\not{\hbox{\kern-4pt $D$}}}
\def\dslash{\not{\hbox{\kern-2pt $\del$}}}

\def\msb{{\bar{\ssstyle M \kern -1pt S}}}

\newcommand{\KaTie}{{\sc Ka\nolinebreak\hspace{-0.3ex}Tie}}

\begin{document}
\begin{titlepage}
\pubblock

\vfill
\Title{New theoretical results\\ in ultrarelativistic
	ultraperipheral lead-lead collisions}
\vfill
\Author{Mariola K{\l}usek-Gawenda, Antoni Szczurek\footnote{Faculty of Mathematics and Natural Sciences, University of Rzesz{\'o}w, ul. Pigonia 1, 35-310 Rzesz{\'o}w}}
\Address{\napoli}
\vfill
\begin{Abstract}
We study dilepton, proton-antiproton and diphoton production in ultraperipheral 
lead-lead collisions at $\sqrt{s_{\mathrm{NN}}}$ = 5.02 and 5.5 TeV.
The nuclear calculations are based on equivalent photon approximation 
in the impact parameter space.
For correct description of the $\gamma\gamma \to p\bar{p} $ Belle data
we include the proton-exchange, the $f_2(1270)$ and $f_2(1950)$ 
s-channel exchanges, as well as the handbag mechanism. 
For four muon production, we take into account electromagnetic (two-photon) 
double-scattering production and direct $\gamma\gamma$ production of four muons in one scattering. 
The cross sections for elementary $\gamma\gamma \to \gamma\gamma$ subprocess 
are calculated including three different mechanisms: 
box diagrams with leptons and quarks in the loops, 
a VDM-Regge contribution with virtual intermediate hadronic excitations 
of the photons and the two-gluon exchange contribution. 
We find that the cross section for elastic $\gamma\gamma$ scattering
could be measured in the lead-lead collisions for the diphoton invariant
mass up to $W_{\gamma\gamma} \approx$ 15 - 20 GeV. 
Our Standard Model predictions are compared with a recent ATLAS
experimental result.

We discuss results for PbPb$\to$PbPb$\mu^+\mu^-\mu^+\mu^-$, 
PbPb$\to$PbPb$e^+e^-e^+e^-$, 
PbPb$\to$PbPb$p\bar{p}$ and PbPb$\to$PbPb$\gamma\gamma$ 
reactions at LHC energy. 
\end{Abstract}
\vfill
\begin{Presented}
EDS Blois 2017, Prague, \\ Czech Republic, June 26-30, 2017
\end{Presented}
\vfill
\end{titlepage}
\def\thefootnote{\fnsymbol{footnote}}
\setcounter{footnote}{0}

\section{Introduction}

Ultraperipheral heavy ion scattering is a special category of nuclear collisions \cite{Budnev,Bertulani}.
This field got a new impulse with the start of the Large Hadron Collider. 
Several final states are possible in ultraperipheral collisions (UPC) of heavy ion. 
The present experiments on UPC concentrated on one- 
\cite{ALICE_rho0,ALICE_Jpsi,ALICE_electrons,CMS_Jpsi}
and two-body final states \cite{ALICE_electrons,ATLAS_photons}. 
Recently we have studied theoretically production in nuclear collisions of one pair of electrons 
and double scattering production of two positron-electron pairs \cite{KGSz_electrons}
as well as single scattering and double scattering production of two $\mu^+\mu^-$ pairs \cite{HKGSz},  
proton-antiproton production \cite{KGLNSz} and elastic $\gamma\gamma$ scattering \cite{KGLSz_photon,KGSSz_photon}.

\section{Theory background}

\begin{figure}[!h]
	\centering
	\includegraphics[scale=0.3]{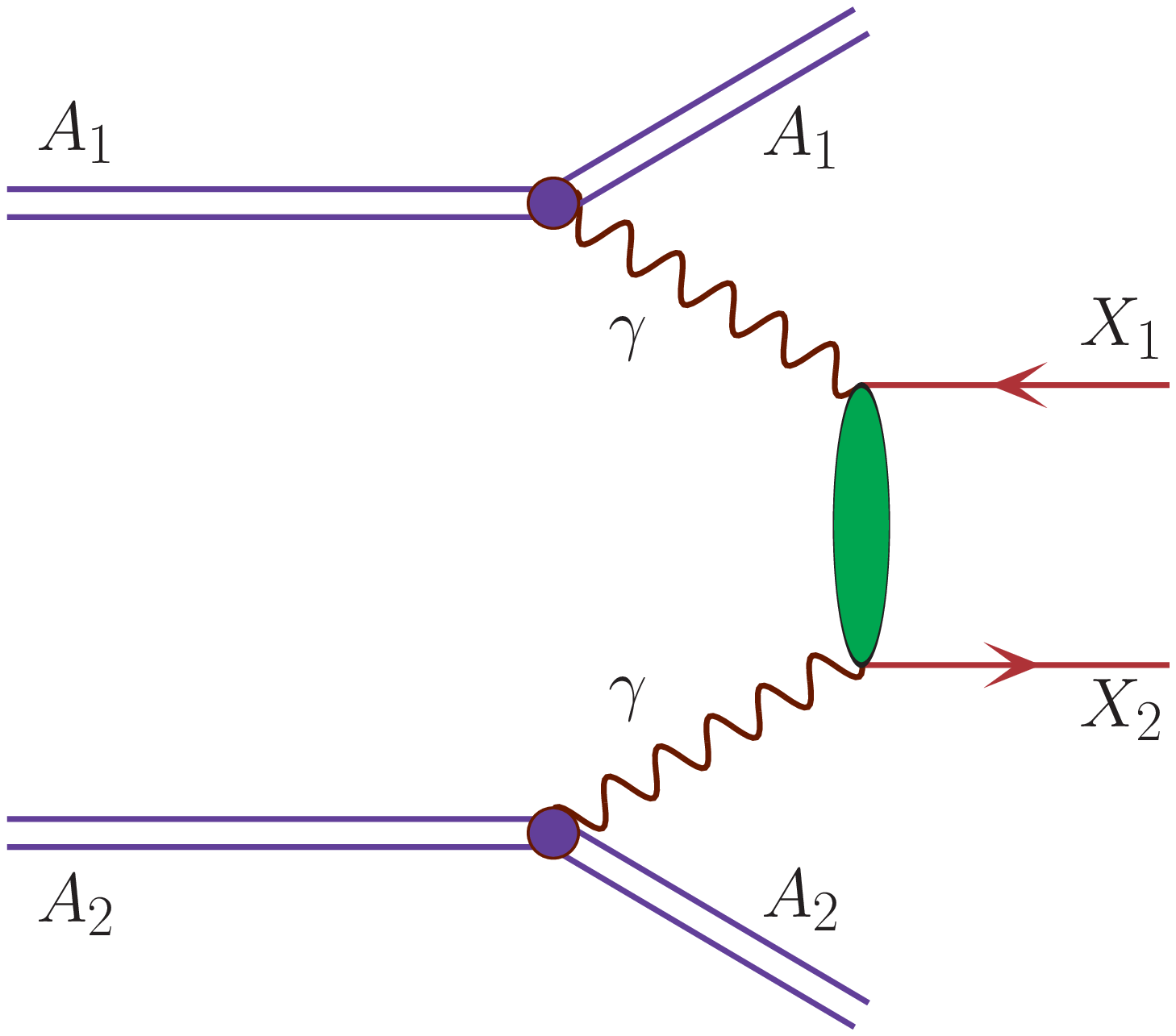}
	\hspace{1.cm}
	\includegraphics[scale=0.3]{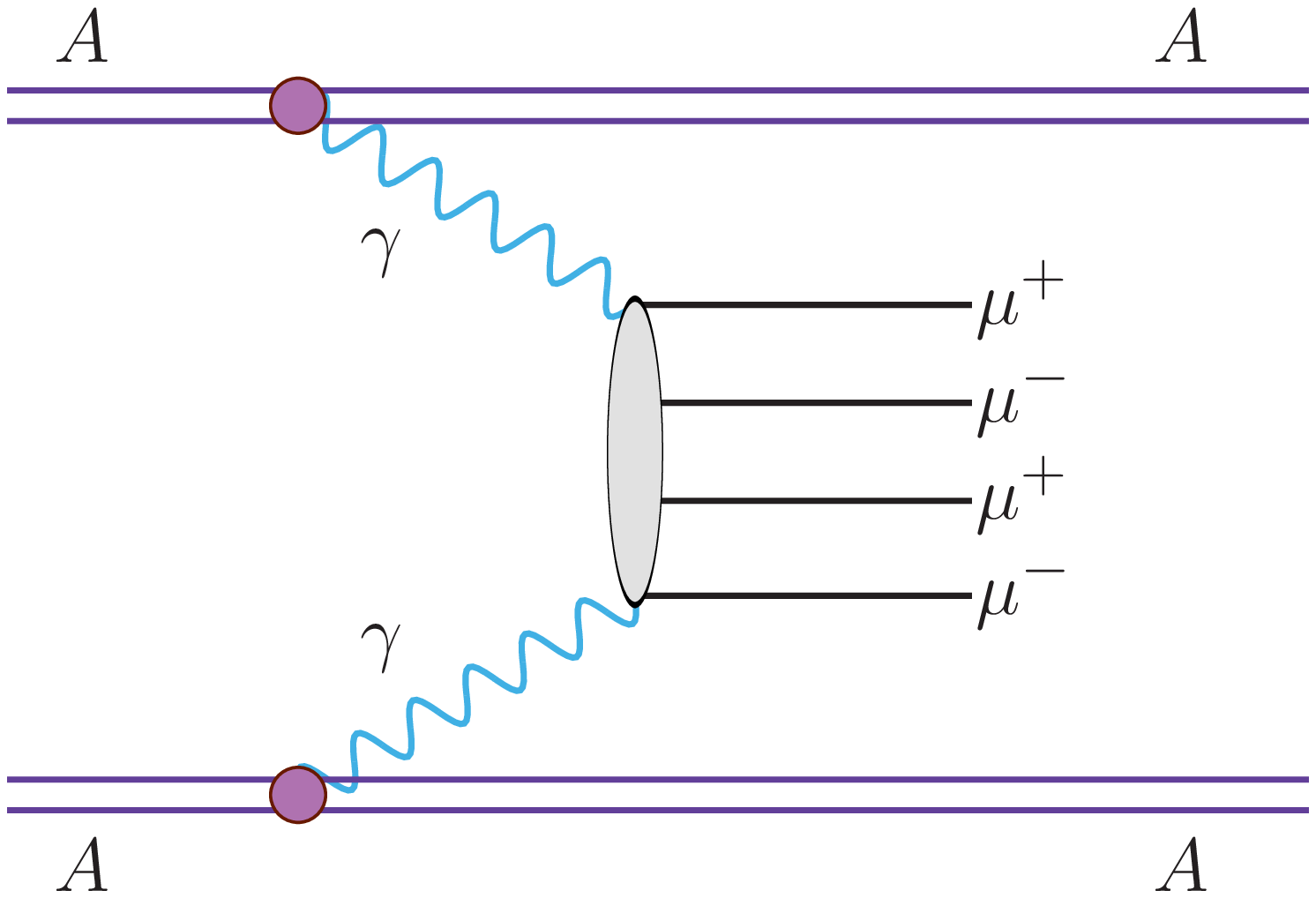} 
		\caption{\label{fig:SS}
		Single-scattering production of a pair of particles (left panel) and
		two pairs of charged leptons (right panel) in ultrarelativistic UPC of heavy ions.}
	\end{figure}
	
\begin{figure}[!h]
	\centering
	\includegraphics[scale=0.3]{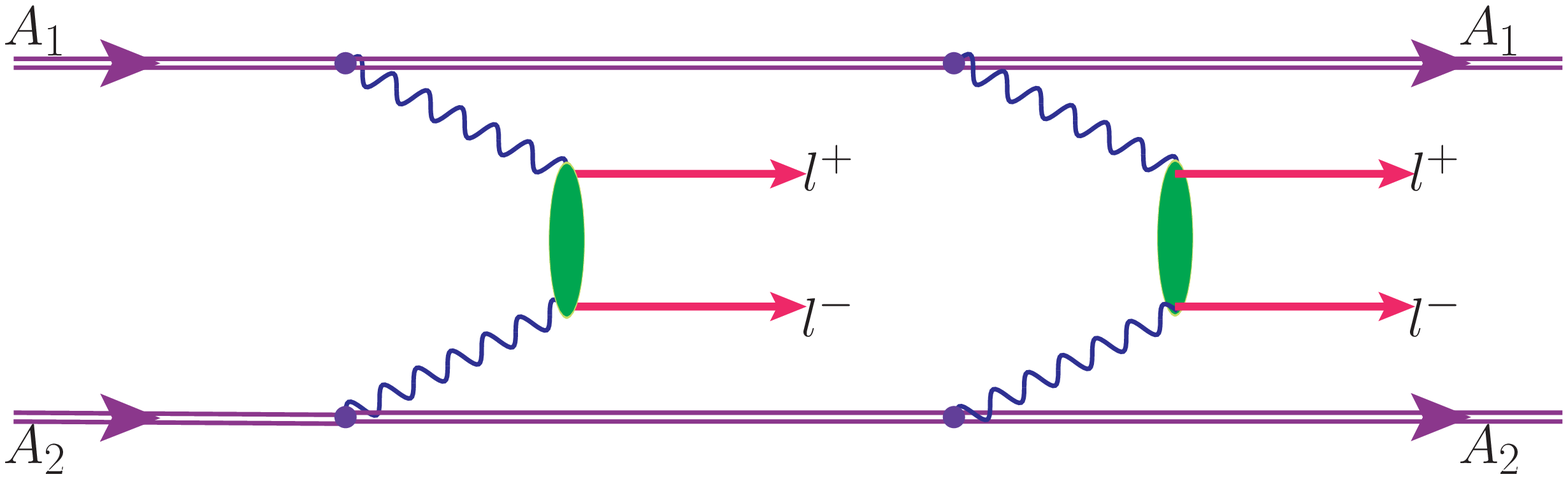}	
	\caption{\label{fig:DS}
		Double-scattering production of two pairs of leptons 			
		in ultraperipheral collisions of heavy ions.}
\end{figure}

A nuclear cross section is calculated in the
equivalent photon approximation in the impact parameter space. 
The total cross section for the single-scattering production (see Fig.~\ref{fig:SS}) 
is expressed through the five-fold integral (for more details see e.g.~\cite{KGSz_muons})
\begin{eqnarray}
\sigma_{A_1 A_2 \to A_1 A_2 X_1 X_2}\left(\sqrt{s_{A_1A_2}} \right) &=&
\int \sigma_{\gamma \gamma \to X_1 X_2} 
\left(W_{\gamma\gamma} \right)
N\left(\omega_1, {\bf b_1} \right)
N\left(\omega_2, {\bf b_2} \right)  \, S_{abs}^2\left({\bf b}\right) \nonumber  \\ 
& \times &
\mathrm{d}^2 b \, \mathrm{d}\overline{b}_x \, \mathrm{d}\overline{b}_y \, 
\frac{W_{\gamma\gamma}}{2}
\mathrm{d} W_{\gamma\gamma} \, \mathrm{d} Y_{X_1 X_2} \;,
\label{eq:EPA_sigma_final_5int}
\end{eqnarray}
where $X_1X_2$ is a pair of produced particles (we considered $p\bar{p}$, $\gamma\gamma$, $e^+e^-$ or $\mu^+\mu^-$;
see left panel of Fig.~\ref{fig:SS}), 
$W_{\gamma\gamma}$ %=\sqrt{4\omega_1\omega_2}$
and $Y_{X_1 X_2}$ %=\left( y_{\gamma_1} + y_{\gamma_2} \right)/2$ 
is invariant mass and rapidity of the outgoing $X_1 X_2$ system. 
Energy of photons is expressed through 
$\omega_{1/2} = W_{\gamma\gamma}/2 \exp(\pm Y_{X_1 X_2})$.
$\bf b_1$ and $\bf b_2$ are impact parameters 
of the photon-photon collision point with respect to parent
nuclei 1 and 2, respectively, 
and ${\bf b} = {\bf b_1} - {\bf b_2}$ is the standard impact parameter 
for the $A_1 A_2$ collision.
Absorption factor $S_{abs}^2\left({\bf b}\right)$ assures that we consider only
ultraperipheral collisions, when the nuclei do not undergo nuclear breakup.
The photon flux ($N(\omega,b)$) is expressed through a nuclear charge
form factor of the nucleus. In our calculations we use a realistic 
form factor which is a Fourier transform of the charge distribution 
in the nucleus.
More details can be found e.g. in \cite{KGSz_muons}.
In our study we calculate also distributions in kinematical variables of each of
the produced particles. Then one can impose easily
experimental cuts on (pseudo)rapidities and transverse momenta of the particles.

The elementary cross section for $\gamma\gamma \to \mu^+\mu^-\mu^+\mu^-$
was calculated with the help of
\KaTie~\cite{Katie}. It is an event generator that is
designed to deal with initial states that have an explicit
transverse momentum dependence, but can also deal with on-shell initial
states, like for this subprocess. \KaTie\ is a parton-level
generator for hadron scattering. A detailed consideration of this process
was studied in Ref.~\cite{HKGSz}.

The cross section for double-scattering production of four leptons (see Fig.~\ref{fig:DS}) is
expressed through a probability density to produce a first and second pair of leptons.
The expression for probability density takes almost the same form as Eq.~(\ref{eq:EPA_sigma_final_5int}).
More details one can find in \cite{KGSz_electrons} and \cite{HKGSz}.

\section{Results}

Before calculating cross section for $l^+l^-l^+l^-$ production ($l=e,\mu$) we have checked 
whether our approach describes the production of a single $l^+l^-$ pair.
In the case of electron-positron production, we get a good agreement 
with the ALICE invariant mass distribution \cite{ALICE_electrons}.
Even imposing experimental cuts relevant for different experiments 
we have obtained cross sections that could be measured
at the LHC even with relatively low luminosity required for UPC of heavy ions of the order of 1 nb$^{-1}$. 
For $L_{int}$ = 1 nb$^{-1}$ and for the main ATLAS detector angular coverage and transverse
momentum cut on each electron/positron $p_t >$ 0.5 GeV we predict 235 events.

Similarly as for the $e^+e^-$ production,  
we get very good description of data for one muon pair production.
We have compared our result with ATLAS preliminary data \cite{ATLAS_muons}.
For a first time, we have shown explicitly that the cross section 
for the single-scattering mechanism is considerably
smaller than the cross section for the double-scattering mechanism. This shows that the
double-scattering mechanism is sufficient for detailed studies and planning experiments.
The cross sections for four-muon production strongly depend on the cuts on muon transverse momenta.

We have discussed in detail also the production of proton-antiproton pairs in photon-photon
collisions \cite{KGLNSz}.
We have shown that the Belle data \cite{Belle} for low photon-photon energies can be nicely described
by including in addition to the proton exchange also the s-channel exchange of the
$f_2(1950)$ resonance, which was observed to decay into the $\gamma\gamma$ and $p\bar{p}$ channels.
Having described the Belle data we have used the elementary $\gamma\gamma \to p\bar{p}$ cross section
to calculate a cross section for $p\bar{p}$ pairs production in heavy ion UPC. 
The integrated (no cuts) cross section for the full phase space is by a factor 5 larger 
than the one corresponding to the Belle angular coverage. 
The larger the range of phase space the broader is the distribution in $y_{diff}=y_p - y_{\bar{p}}$.
In Ref.~\cite{KGLNSz} we have shown predictions for experimental
cuts corresponding to the ATLAS, CMS and ALICE experiments. We find cross
sections of 35 $\mu$b for the ALICE cuts ($|y| <$ 0.9, $p_t >$ 1 GeV) 
and 155 $\mu$b taking into account ATLAS or CMS cuts ($|y| <$ 2.5, $p_t >$ 0.5 GeV).

In addition we have studied how to measure elastic photon-photon 
scattering in lead-lead UPC \cite{KGLSz_photon,KGSSz_photon}. 
The cross section for photon-photon scattering was calculated taking 
into account well known box diagrams with elementary standard 
model particles (leptons and quarks), 
a VDM-Regge component as well as a two-gluon exchange mechanism,
including massive quarks, all helicity configurations of photons and massive and massless gluons. 
For the PbPb$\to$PbPb$\gamma\gamma$ reaction 
we identified regions of the phase space where the two-gluon contribution 
could be enhanced relatively to the box contribution. 
The region of large rapidity difference between the two emitted photons 
and intermediate transverse momenta 1 GeV  $<p_t<$  2 - 5 GeV seems
optimal in this respect.

This year the ATLAS Collaboration published a new result \cite{ATLAS_photons} 
for light-by-light scattering in quasi-real photon interactions 
in ultraperipheral lead-lead collisions at 
$\sqrt{s_{\mathrm{NN}}}$= 5.02~TeV. 
The measured fiducial cross section which includes limitation 
on photon transverse momentum, photon pseudorapidity, diphoton invariant mass, 
diphoton transverse momentum and diphoton accoplanarity, 
were measured to be 70 $\pm$ 20 (stat.) $\pm$ 17 (syst.) nb. 
This result is compatible with the value of 49 $\pm$ 10 nb predicted by us for
the ATLAS cuts and experimental luminosity.

%\section{Conclusion}

%We use the Weizs{\"a}cker-Williams approach which is a semiclassical approximation
%scheme used to analyze electromagnetic intaractions. 
%The strong electromagnetic field of ultrarelativistic nuclei is a source of intense photon fluxes
%which can provide a possibility to study two-photon interaction in UPC. 
%This study was focused only on ultraperipheral collisions of heavy ions.
%This means that distance between nuclei was bigger than sum over radii of two nuclei.
Over the past few years, many exciting UPC results have been presented.
We hope that our new theoretical predictions will be a source of inspiration
for future experiments.

%%%%%%%%%%%%%%%%%%%%%%%%%%%%%%%%%%%%%%%%%%%%%%%%%%%%%%%%%%%%%%%%%%%%%%%%%
%%
%%   use this format to include an .eps figure into your paper
%%
% \begin{figure}[htb]
% \centering
% \includegraphics[height=1.5in]{magnet}
% \caption{Plan of the magnet used in the mesmeric studies.}
% \label{fig:youfigure}
% \end{figure}
%%%%%%%%%%%%%%%%%%%%%%%%%%%%%%%%%%%%%%%%%%%%%%%%%%%%%%%%%%%%%%%%%%%%%%%%%%%

%%%%%%%%%%%%%%%%%%%%%%%%%%%%%%%%%%%%%%%%%%%%%%%%%%%%%%%%%%%%%%%%%%%%%%%%%
%%
%%   use this format to include a LaTeX table  into your paper
%%
% \begin{table}[t]
% \begin{center}
% \begin{tabular}{ccc}  
%  ...
% \end{tabular}
% \caption{Blood cyanide levels for the two patients.}
% \label{tab:blood}
% \end{center}
% \end{table}
%%%%%%%%%%%%%%%%%%%%%%%%%%%%%%%%%%%%%%%%%%%%%%%%%%%%%%%%%%%%%%%%%%%%%%%%%%%

\Acknowledgements % if needed
This work was partially supported by the Polish National Science Centre grant DEC-2014/15/B/ST2/02528.

\end{document}